\documentclass[pmlr,twocolumn,10pt]{jmlr} 

\usepackage{xcolor}
\usepackage{booktabs}

\usepackage[load-configurations=version-1]{siunitx} 
\DeclareMathOperator*{\argmin}{arg\,min}

\DeclareMathOperator*{\supp}{supp}
\newcommand{\cossim}[2]{\ensuremath{\lvert \cos\angle(#1,#2)\rvert}}

\theorembodyfont{\upshape}
\theoremheaderfont{\scshape}
\theorempostheader{:}
\theoremsep{\newline}

\jmlrvolume{LEAVE UNSET}
\jmlryear{2021}
\jmlrsubmitted{LEAVE UNSET}
\jmlrpublished{LEAVE UNSET}
\jmlrworkshop{Machine Learning for Health (ML4H) 2021} 

\title[]{Exploring latent networks in resting-state fMRI using voxel-to-voxel causal modeling feature selection}

\author{%
\Name{Hassan Baker} \Email{bakerh@udel.edu}\and
\Name{Austin J. Brockmeier} \Email{ajbrock@udel.edu}\\
\addr Department of Electrical and Computer Engineering, University of Delaware, Newark, Delaware, USA
}

\begin{document}

\maketitle
\begin{abstract}
Functional networks characterize the coordinated neural activity observed by functional neuroimaging. The prevalence of different networks during resting state periods provide useful features for predicting the trajectory of neurodegenerative diseases. Techniques for network estimation rely on statistical correlation or dependence between voxels. Due to the large number of voxels, rather than consider the voxel-to-voxel correlations between all voxels, a small set of seed voxels are chosen. Consequently, the network identification may depend on the selected seeds. As an alternative, we propose to fit first-order linear models with sparse priors on the coefficients to model activity across the entire set of cortical grey matter voxels as a linear combination of a smaller subset of voxels. We propose a two-stage algorithm for voxel subset selection that uses different sparsity-inducing regularization approaches to identify subject-specific causally predictive voxels. To reveal the functional networks among these voxels, we then apply independent component analysis (ICA) to model these voxels' signals as a mixture of latent sources each defining a functional network. Based on the inter-subject similarity of the sources' spatial patterns we identify independent sources that are well-matched across subjects but fail to match the independent sources from a group-based ICA. These are resting state networks, common across subjects that group ICA does not reveal. These complementary networks could help to better identify neurodegeneration, a task left for future work. 
\end{abstract}

\section{Introduction}
\label{sec:intro}
Functional magnetic resonance imaging (fMRI) signal is among the most popular techniques for identifying functional connections (FC) across regions of the brain. Resting-state fMRI (rs-fMRI) is a task-free neuroimaging paradigm. It reveals the neural activities while the subject is not doing anything in particular. The importance of studying rs-fMRI lies in its ability to reveal the hidden functional relationships among  brain regions, providing  baselines or prototypes of the expected neural activities for healthy subjects versus unhealthy subjects who suffer from neurodegenerative diseases such as schizophrenia \citep{karbasforoushan2012resting, baker2014disruption} and Alzheimer's disease \citep{de2018comprehensive}. A better understanding of rs-fMRI may lead to an early-stage detection for neural diseases and also other kinds of diseases \cite{kucikova2021resting}. 

Often fMRI processing techniques focus on the statistical correlation of the fMRI time series or are based on seed selection. The latter can create variation among the selected seeds \citep{lv2018resting}. An alternative approach is to try to model the causal relationship (in the sense of Granger) among voxels directly. There are several works of modeling brain activities using causal structure. \cite{roebroeck2005mapping} used causality for identifying the influence of any selected regions of interest during complex visuomotor tasks. \cite{liao2010evaluating}  were able to find specific causal influences among the detected resting-state networks using causal structure applied on rs-fMRI. \cite{liao2011small} applied multivariate Granger causality analysis and graph theory on rs-fMRI to identify pivotal hubs and their interactions with each other. Their work relied on using the pooled activity of voxels in predefined regions of interest. \cite{dsouza2017exploring} explored causality structure on resting-state data, using principal component analysis (PCA) to overcome the problem of underdeterminedness. 
\cite{hamilton2011investigating} and \cite{gao2020altered} were able to identify unique neural patterns for major depressive disorder and increases and decreases of Granger causality within different brain regions for Schizophrenic patients.

In this paper, we focus on overcoming the limitations of other techniques mentioned above. In particular, we provide a voxel-level methodology that enables whole-brain voxel subset selection, without either dimensionality reduction, seed selection, or pooling within regions of interest.  Firstly, we focus on the voxel subset selection procedure based on the ability of the voxels to represent the dynamics in the brain under a sparsity assumption. We then perform independent component analysis (ICA) on the selected voxels. ICA optimizes a demixing matrix to linearly combine these voxels revealing the underlying  sources that account for higher-order dependency among the selected voxels. 
\section{Data and Methodology}
The resting-state fMRI data is obtained from \citet{risk2021multiband}. We choose the dataset with the highest multiband (i.e., MB12) as it gives the highest temporal resolution at $\sim$2~Hz. A subset of 24 healthy subjects are chosen and their preprocessing is done by using \texttt{fmriprep}, 6 subjects are held-out for future analysis. The rest of the analysis is done on the grey-matter cortical voxels. To evaluate the causal models, the fMRI time series (688 time points corresponding to $\sim$352~s) for each subject is divided with roughly the first 80\% (550 points) for training, the next 10\% (68 points) for validation, and the final 10\% (70 points) for testing. 

Let $\mathcal{S}$ denote the set of $\lvert \mathcal{S} \rvert$ subjects, where $\lvert\cdot\rvert$ denotes the cardinality of the set. For $s \in \mathcal{S}$, the fMRI training data forms a matrix, $X^{(s)}  \in \mathbb{R}^{V^{(s)}\times T}$, $V^{(s)}$ is the number of cortical grey voxels per subject and $T$ is the number of time points.\footnote{~\tableref{tab:grey_voxels} in the Appendix shows the number of cortical grey matter voxels per subject.} When the modeling is applied to any subject we will drop the subject indices and refer to $X^{(s)}$ and $V^{(s)}$ by $X$ and $V$, respectively. 

\subsection{Voxel Subset Selection}
The goal of voxel subset selection is to find a subset of voxels that are predictive of the fMRI signal in the cortical grey matter voxels for a given subject. A key motivation for it is to overcome the computational complexity when performing a voxel-to-voxel analysis that considers the temporal dynamics of a large number of grey-matter voxels. Due to the number of voxels compared with the number of time frames, robust voxel subset selection is non-trivial. 

A two-stage approach that combines a fine-grained region based atlas and sparse linear regression is used to find a predictive subset.\footnote{Fig.~\ref{fig:explanation} in the Appendix shows a workflow diagram that summarizes the approach.} 

In the first stage, the 1000 region Schaeffer atlas~\citep{schaefer2018local} is used to organize the cortical regions of the brain into different groups based on their functional connectivity (FC). The 1000 region atlas is spatially fine-grained which helps selecting representative voxels from different regions of the cortical region. Let $R_i \in  \{1,\ldots, V\}^{V_i} \subset \mathbb{N}^{V_i}, i \in \{1,\dots,1000\}$ be a region represented as a vector of voxel indices associated to the $i$-th region of the atlas, and $V_i$ is the number of voxels in the region. We identify a subset of voxels within each region by their ability to predict the activities of the rest of the cortical region. A single-lag predictive model is used $ X_{t}^{\neg  i}   \approx  W_{R_i} X_{t-1}^{i}$, where $X_{t}^{\neg  i}  \in \mathbb{R}^{(V - V_i) \times T-1}$ is a time series starting at the second time index of all voxels except the voxels from the region $R_i$, $X_{t-1}^{i}  \in \mathbb{R}^{V_i \times T -1}$ is the matrix of delayed time series that starts from the first time index and ends at the second to last for voxels within region $R_i$, and $W_{R_i} \in \mathbb{R}^{(V-V_i) \times V_i}$ is the mapping matrix of the previous neural activities in region $R_i$ to the next neural activities in the rest of the cortical regions. To summarize, we develop predictive models that predict the activity in other cortical regions based on the activity in region $R_i$ at the previous time point.

We assume that $W_{R_i}$ has a special sparsity structure: most of the columns are zeros as they do not contribute to the predictive model. Thus, to estimate the matrix $W_{R_i}$ we add an $\ell_{2,1}$ penalty to the sum of squared errors,
\begin{align*}
   W^*_{R_i} =  \argmin_{W \in \mathbb{R}^{(V-V_i) \times V_i}  } \| X_{t}^{\neg  i}  - W X_{t-1}^{i} \|_F^2 + \lambda_{2,1}\|W\|_{2,1},  \label{eq:selection_stage_1}
\end{align*}
where $\| \cdot \|_F^2$ denotes the Frobenius norm and $\|W\|_{2,1} = \sum_j \sqrt{ \sum_i W_{ij}^2}$. The solution to this optimization problem is obtained using the method detailed by \cite{nie2010efficient}. The selection of the regularization parameter $\lambda_{2,1}$ is done by performing the analysis over 20 equally line spaced points in the interval $[\lambda_{21}^\mathrm{max}$, .001$\lambda_{21}^\mathrm{max}]$, where $\lambda_{21}^\mathrm{max}$ is the regularization that returns a zero solution. After fitting the model, an unshrunk model is obtained by fitting an ordinary least squares model on the nonzero columns. The parameter $\lambda_{21}$ that achieves the minimum validation mean squared error (based on the unshrunk ordinary least square model coefficients) is chosen.   

This voxel selection procedure aims to select voxels within a group. Let  $\supp(W_{R_i}^*)$ be the indices of non-zero columns of $W_{R_i}^*$.  Abusing vector indexing notation, the corresponding set of voxel indices for region $i$ are $R_i(\supp(W_{R_i}^*))$. Let $\mathcal{V}_{\text{S1}}= R_1(\supp( W_{R_1}^*)) \cup \cdots \cup R_2(\supp( W_{R_{1000}}^*))$ be the set of selected voxels from the first stage with $V_{\text{S1}} = \lvert  \mathcal{V}_{\text{S1}} \rvert $.

In the second stage, we assume there  are voxels across regions that are redundant or unnecessary for predicting any given voxel. Hence, we solve the following LASSO model to choose a subset of the previously selected voxels to explain each voxel, $j \in \{1,\dots,V\}$, independently:
\begin{align*}
     \mathbf{w}^{**}_j = \argmin_{\mathbf{w} \in \mathbb{R}^{V_{\text{S1}}}} \| \mathbf{x}_{t}^j -   (X_{t-1}^{\text{S1}})^\top \mathbf{w} \|_2^2 + \lambda_j\|\mathbf{w}\|_{1},
\end{align*}
where $\mathbf{x}_{t}^j \in \mathbb{R}^{T-1}$ is the fMRI time series for voxel $j$, $X_{t-1}^{\text{S1}} = ([\mathbf{x}_{t-1}^k]_{k\in \mathcal{V}_{\text{S1}}})^\top  \in \mathbb{R}^{V_{\text{S1}} \times T-1}$ is the selected voxels time series from the first stage, and $W^{**}=[\mathbf{w}^{**}_1, \ldots , \mathbf{w}^{**}_V] \in \mathbb{R}^{V_{\text{S1}} \times V  }$ is the mapping matrix from the previous activities of the first-stage selected voxels to all grey matter cortical voxels. 
The same procedure of the first stage is applied here to select the regularization parameter $\lambda_j$, except that only 10 linearly spaced values in the range $[\lambda_{j}^\mathrm{max}, 10^{-3}\lambda_{j}^\mathrm{max}]$ are tested. Note that each voxel $j\in\{1,\ldots,V\}$ has its own regularization parameter chosen from a voxel-dependent range. The subset of selected voxels from the second stage is $\mathcal{V} = \cup_{j=1}^{V} \mathbf{v}_\text{S1}(\supp(\mathbf{w}_j^{**}))$, where $\mathbf{v}_\text{S1} = [ k]_{k\in \mathcal{V}_\text{S1}}$.

The subsequent analysis is based on the selected voxels from the second stage along with coefficients from a final ridge regression model $W^{***} = [ \mathbf{w}_1^{***},\ldots,\mathbf{w}_V^{***}  ]  \in \mathbb{R}^{V_{\text{S1}} \times V}$ is the mapping matrix across all voxels, where 
\begin{align*}
     \mathbf{w}_j^{***} = \!\!\!\!\argmin_{\substack{\mathbf{w} \in \mathbb{R}^{V_{\text{S1}}}\\ \supp(\mathbf{w})\subseteq \supp( \mathbf{w}^{**}_j)} } \!\!\!\! \| \mathbf{x}_{t}^j -  (X_{t-1}^\text{S1})^\top  \mathbf{w}\|_2^2 + \mu_j\| \mathbf{w} \|_{2}^2,  
\end{align*}
and $\mu_j$ is selected by the validation mean squared error (MSE). To test the significance of the causality, models are built using 100 time-shuffled versions of the training data, and  test set MSEs are obtained. Averaged across subjects, 51\% of the cortical grey matter voxels have testing MSEs statistically lower then the MSEs of the shuffled models (p-value $\le$ 0.05). The range across subjects is 38\%--75\%.

\subsection{Analysis of Selected Voxels}

Although the selected voxels are assumed to capture the dynamics in the brain, it is not guaranteed that these voxels are statistically independent.  Thus, we apply ICA to organize the selected voxels into latent/hidden networks per subject. This also enables a direct comparison with group-based ICA, which combines the data from all subjects assuming common networks across subjects. 

As an alternative, we apply ICA (in particular, fastICA, with the $\mathrm{logcosh}$ contrast function) directly on the matrix of time-series from the subset of predictive voxels: $(X^\text{S2})^\top =[\mathbf{x}^j]_{j\in\mathcal{V} }  = M S \in \mathbb{R}^{T \times |\mathcal{V}| }$, where $M \in \mathbb{R}^{T\times \lvert\mathit{IC}\rvert}$ is the mixing matrix, $\lvert\mathit{IC}\rvert$ is the number of components, $S \in \mathbb{R}^{\lvert\mathit{IC}\rvert \times |\mathcal{V}|}$ is the matrix of independent components (ICs). Using the causal model coefficients $W^{***}$, all voxels in the grey matter cortical regions are approximated as a mixture of these ICs, $X \approx W^{***} S^\top M^\top = Q M^\top$. For subject  $s \in S$, let $Q^{(s)} \in \mathbb{R}^{V^{(s)}\times \lvert\mathit{IC}\rvert} $ denote the spatial pattern of the ICs across the whole set of grey matter cortical voxels. Since each subject has a different set of cortical grey matter voxels $V^{(s)}$, we still need to define a common space to define the ICs. To do this we project each IC back into the MNI152 space and apply a spatial blur to account for the grey-matter misalignment across subjects, with $\mathit{IC}_k^{(s)} = \mathrm{blur}(\mathrm{map}_\text{MNI152}(Q^{(s)}_k))$ for $k\in\{1,\ldots,\lvert\mathit{IC}\rvert\}$. Blurring uses a 3D Gaussian kernel with $\sigma=3$ voxels. 

 To compare our method, we apply CanICA \citep{varoquaux2010group} which is a popular group-level ICA approach that combines the subjects' fMRI data and extracts the ICs based on this group data in the MNI152 space.  We also apply the Gaussian kernel to each group IC $\mathit{IC}^\textit{group}_k,\quad k\in\{1,\ldots,\lvert\mathit{IC}\rvert\}$. We set the number of IC's to 20 as it is the number used in previous studies \citep{schultz2014template, biswal2010toward} and is also the default number of components in the Group ICA toolbox such as \texttt{GIFT}  \citep{calhoun2001method} and \texttt{CanICA} in \texttt{Nilearn} framework. 

To test the consistency of our approach, we calculate the \textit{inter-subject similarity} (IS) for each IC, $\mathit{IC}_k^{(s)}$, and subject $s'$, which is denoted by $\mathrm{IS}(\mathit{IC}_k^{(s)}, s')$. We define it as the maximum absolute cosine similarity of the $k$-th IC from subject $s$ with each IC from subject. Mathematically,
\begin{align*}
&\mathrm{IS}(\mathit{IC}_k^{(s)}, s') = \max_{k' \in \{1,\dots,\lvert\mathit{IC}\rvert\}} \cossim{\mathit{IC}_k^{(s)}}{\mathit{IC}_{k'}^{(s')}}\\
&\mathrm{IS}(\mathit{IC}_k^{(s)})= [\mathrm{IS}(\mathit{IC}_k^{(s)}, 1),\ldots, \mathrm{IS}(\mathit{IC}_k^{(s)}, |S|)] \in R^{|S|},
\end{align*}
where $\cossim{\cdot}{\cdot}$ is the absolute cosine similarity. Note that when $s=s'$, $\mathrm{IS}(\mathit{IC}_k^{(s)}, s) =1$. 

We also calculate the similarity between the individual ICs and the group ICs, and refer to this as the \textit{Individual-Group Similarity} (IGS):
\begin{align*}
&\mathrm{IGS}(\mathit{IC}^{(s)}_k) = \max_{k' \in \{1,\dots,\lvert\mathit{IC}\rvert\}} \cossim{\mathit{IC}^{(s)}_k}{\mathit{IC}_{k'}^\textit{group}}.
\end{align*}

\vspace{-.5cm}
\section{Results}

We are interested in the ICs that have low $\mathrm{IGS}$ and have high values of $\mathrm{IS}$ as they explain consistent functional networks identified through the proposed approach that are distinct from the group-based ICA. Initially, we tested whether the subject-specific ICs that have at least half of the values in $\mathrm{IS}(\mathit{IC}^{(s)}_k)$ larger than the value of $\mathrm{IGS}(\mathit{IC}^{(s)}_k)$. 

The vast majority of the subjects satisfy this condition (only 2 subjects failed), which means that the majority of similarity values between individual ICs of different subjects are greater than the similarity value with the group ICs. 

We then performed a hierarchical clustering (weighted Euclidean linkage) on the individual ICs. Instead of defining the clusters in terms of the similarity of the ICs directly, we concatenate and compare the $\mathrm{IS}$ and $\mathrm{IGS}$ values, computing a weighted Manhattan distance with a weight of $|S|$ for the $\mathrm{IGS}$ entry and a weight of 1 for the $\mathrm{IS}$ entries.

\begin{figure}[h!bt]
\floatconts
  {fig:nodes}
  {\caption{Hierarchical clustering of subject ICs based on their IS and IGS values. Column labeled 24 is the IGS values.}}
  {\includegraphics[width=1\linewidth]{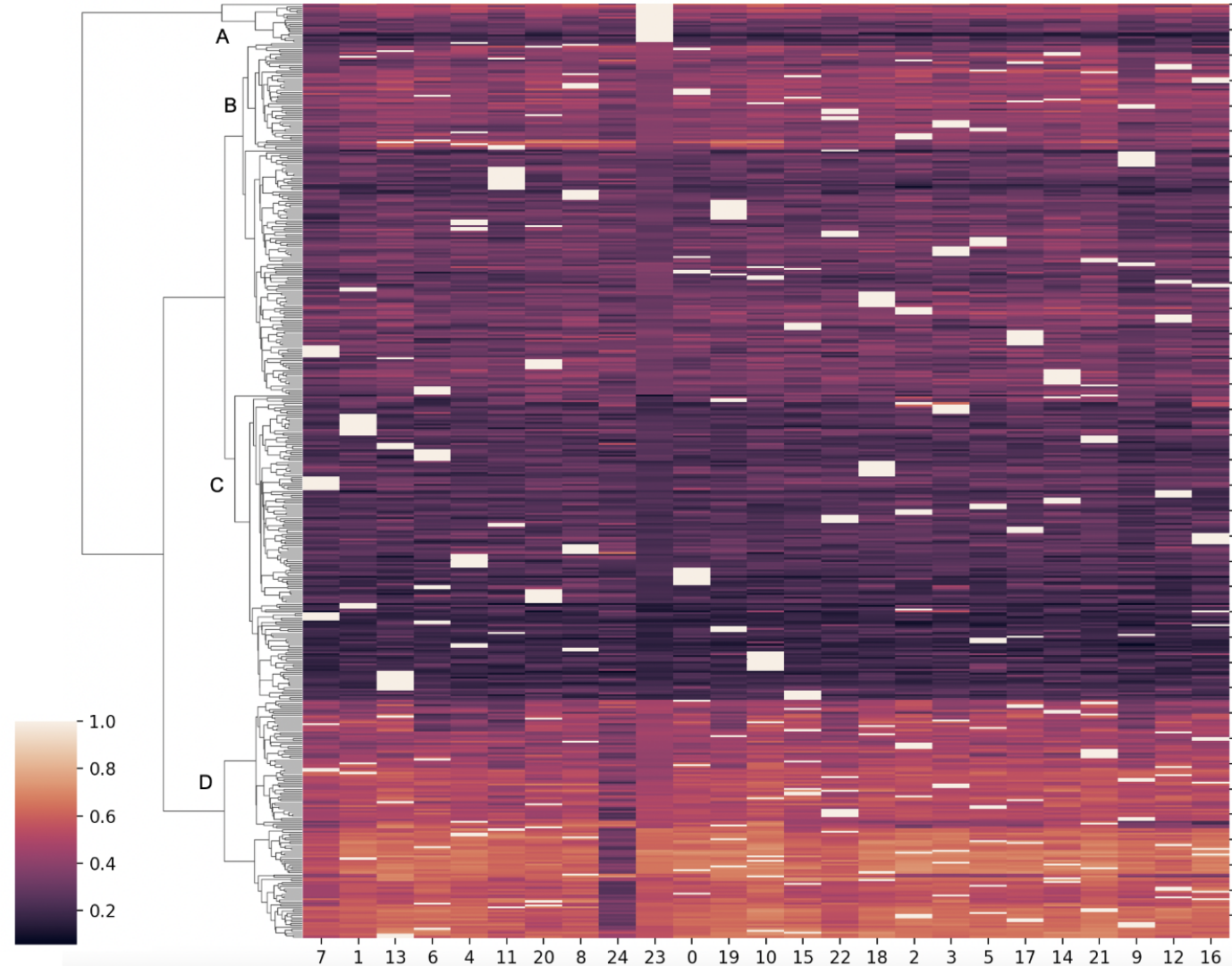}}
\label{fig:cluster}
\end{figure}

Fig.~\ref{fig:cluster} shows the clustering performed on the feature vector. Column labeled 24 is the $\mathrm{IGS}$ values. The dendrogram shows four different clusters: A, B, C, D.  
Cluster D is the set of ICs that have high $\mathrm{IS}$ and low $\mathrm{IGS}$. That is, ICs in cluster D identify consistent spatial patterns of the sources associated to selected voxels but are not similar to any group IC. Notably, all subjects but four subjects have at least one IC in cluster D.

\begin{figure}[h]
\floatconts
  {fig:nodes}
  {\caption{Coronal slices of individual ICs in cluster D (columns 1–4) and group ICs (column 5). }}
  {\includegraphics[width=1\linewidth]{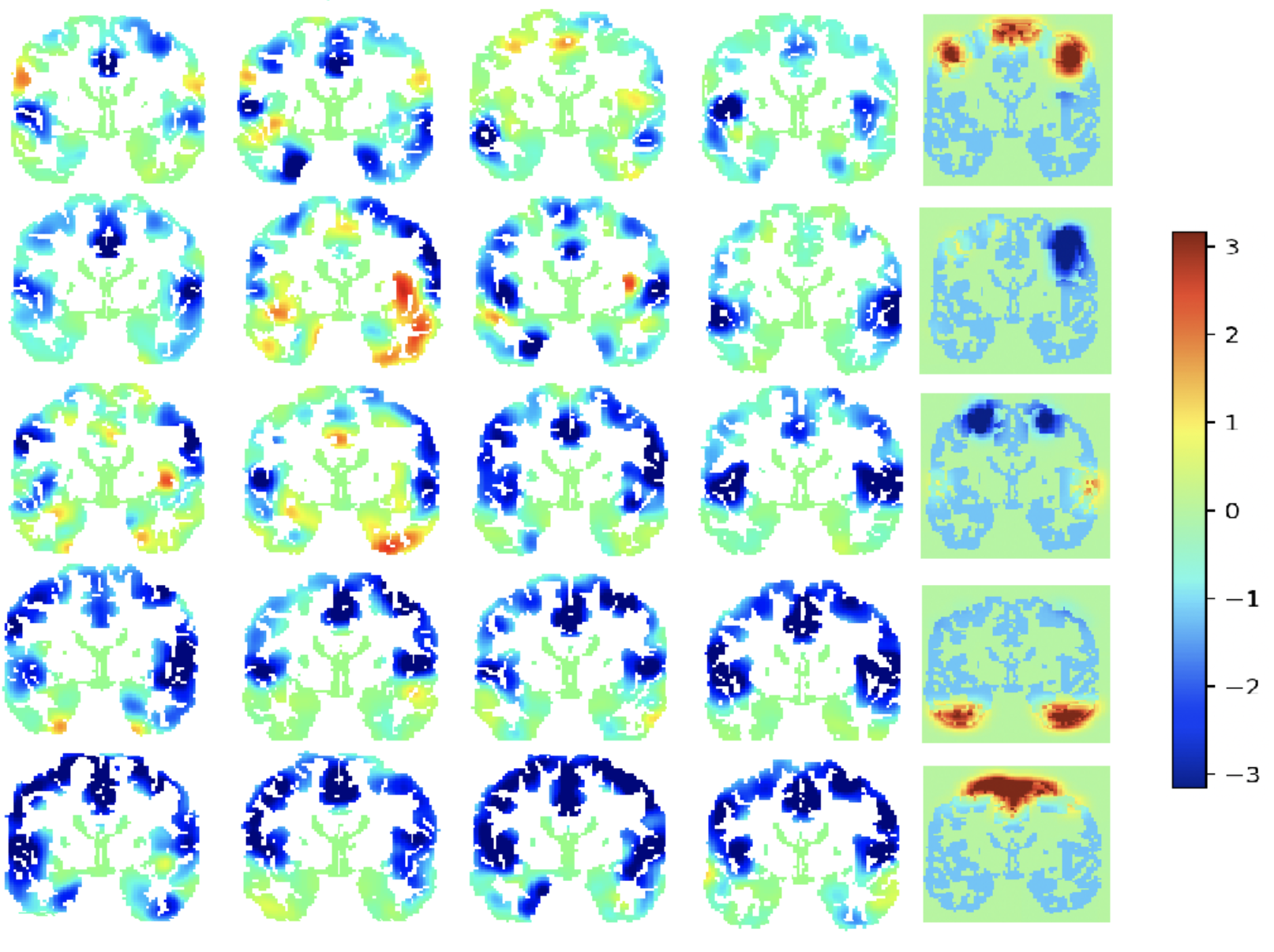}}
\label{fig:ic-across-subjects}
\end{figure}

We focus on cluster D to investigate if there are features that distinguish individual ICs from group ICs. We show a coronal slice from individual ICs belonging to cluster D from different subjects  in Fig.~\ref{fig:ic-across-subjects}. 
Even in these slices the similarities between ICs are evident, which reflects the high $\mathrm{IS}$ values. For contrast, the most similar group ICs are displayed in the last column with green backgrounds. Clearly, the group ICs are much sparser compared with the individual ICs. The individual ICs found through causal modeling may capture more global functional networks rather than the localized functional networks of group ICA.
\section{Conclusion and Future Work}
We introduced an automated and scalable voxel selection method based on first-order linear causal models with sparsity inducing regularizations. We show that our method is able to recover unique spatial patterns across subjects compared to a group-based method. Whether these patterns are meaningful in distinguishing healthy versus non-healthy subjects is left for future work. 

 \acks{Research was carried out with the support of the University of Delaware Research Foundation. This research was supported in part through the use of Data Science Institute (DSI) computational resources at the University of Delaware, specifically the DARWIN system. The DARWIN computing project at the University of Delaware is supported by the National Science Foundation under Grant No.~OAC-1919839. The authors thank the support from the University of Delaware IT Research Computing Group.}
\bibliography{jmlr-sample}


\section{Appendix}\label{apd:first}
\begin{figure*}[htbp]
\floatconts
  {fig:ml4h}
  {\caption{In the first stage, we select the representative voxels of region $R_i$ by estimating an $\ell_{2,1}$-norm regularized linear model that predicts the activities of other regions based on one time step in advance of neural activities of region $R_i$. The representative voxels are the ones which corresponds to the nonzero columns. In the second stage, the goal is to select voxels from the voxels selected from the first stage. We apply a $\ell_1$-norm regularized linear model (LASSO) that predicts the activity of each cortical voxel based on one time step in advance of neural activities of the first stage selected voxels. The final voxels selection is based on the corresponding voxels of nonzero coefficients. In other words, if voxel has at least a nonzero coefficient that is predictive of any voxels activity, then it will be selected. }}
  {\includegraphics[width=\linewidth]{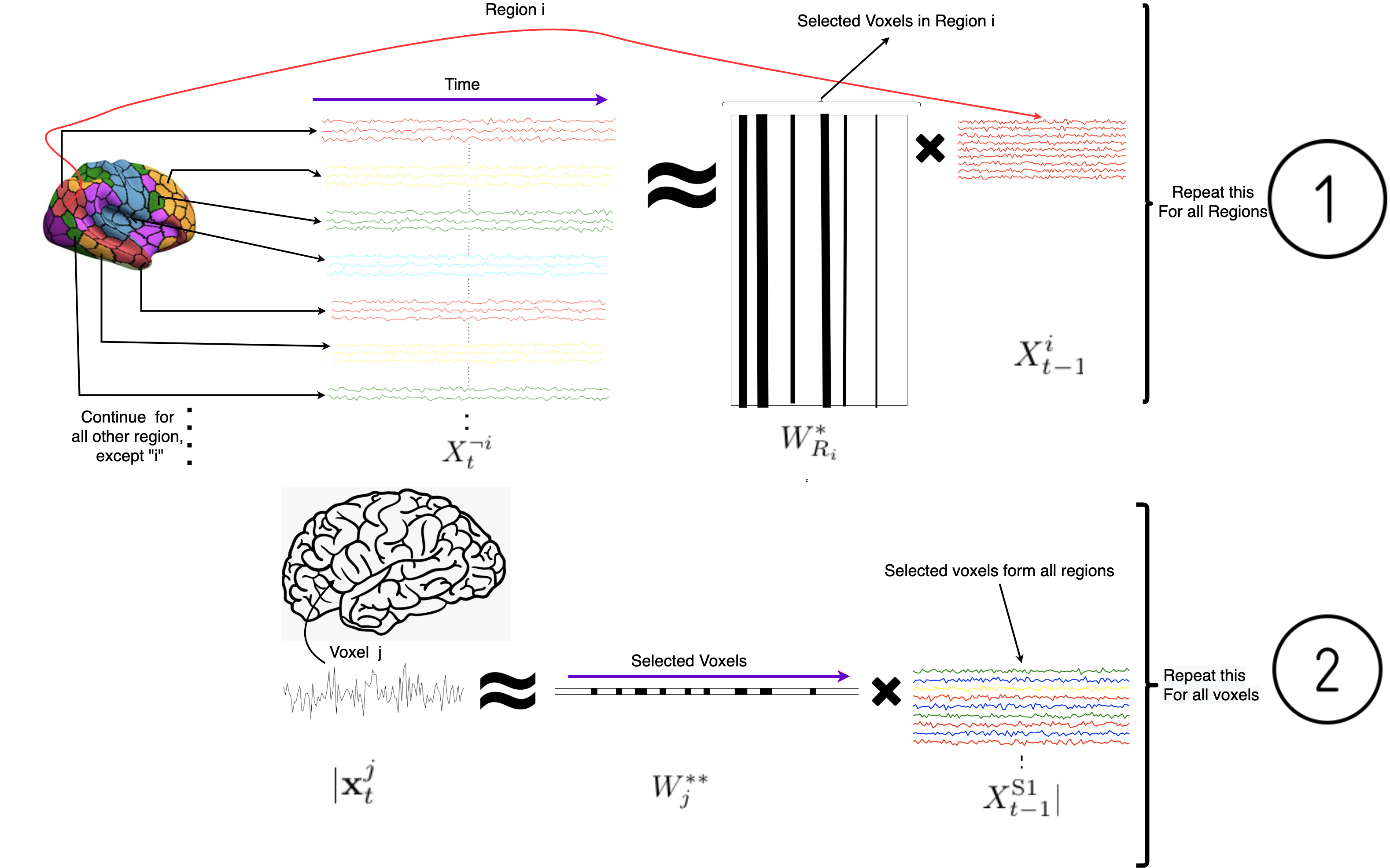}}
\label{fig:explanation}
\end{figure*}

We show the schematic diagram for the two-stage voxel selection procedure in Fig.~\ref{fig:explanation}.  We also show the number of cortical grey matter voxels for each subject in~\tableref{tab:grey_voxels}.

\begin{table}[hbtp]
\floatconts
  {tab:grey_voxels}
  {\caption{The number of cortical grey matter voxels $V^{(s)}$ for each subject.}}
  {\begin{tabular}{rr}
  \toprule
  \bfseries Subject & \bfseries Voxels\\
  \midrule
 02 & 97\,837 \\ 
 03 & 95\,915  \\
 07 & 102\,332 \\
 08 & 99\,330 \\
 10 & 96\,809 \\
 11 & 98\,766 \\
 12 & 92\,915 \\
 13 & 97\,803 \\
 14 & 100\,182 \\
 15 & 96\,888 \\
 16 & 98\,085 \\
 17 & 97\,728  \\
 18 & 97\,810  \\
 19 & 98\,337 \\
 20 & 99\,218 \\
 22 & 98\,011 \\
 23 & 96\,135 \\
 24 & 98\,981 \\
 25 & 95\,911 \\
 26 & 100\,367 \\
 27 & 94\,910 \\
 28 &  98\,893 \\
 29 & 96\,896 \\
 30 & 95\,460 \\
  \bottomrule
  \end{tabular}}
\end{table}

\end{document}